\documentclass{elsart}

\usepackage{graphicx,amssymb}
\journal{pla}

\newcommand{\de}[1]{\left( #1 \right)}
\newcommand{\De}[1]{\left[ #1 \right]}
\newcommand{\DE}[1]{\left\{ #1 \right\}}

\newcommand{\ket}[1]{\left| #1 \right\rangle}
\newcommand{\bra}[1]{\left\langle #1 \right|}

\newcommand{\abs}[1]{\left| #1 \right|}
\newcommand{\mean}[1]{\left\langle #1 \right\rangle}
\newcommand{\op}[1]{{\mathbf{#1}}}

\newcommand{\up}{\uparrow}
\newcommand{\down}{\downarrow}

\newcommand{\ie}{{\it{i.e.}}: }
\newcommand{\ettal}{{\it{et al.}}, }

\newcommand{\pra}[4]{{Phys. Rev. A} {{#2}} (#4) #3}

\newcommand{\prl}[4]{{\it{Phys. Rev. Lett.}} {{#2}} (#4) #3}

\newcommand{\pla}[4]{{{Phys. Lett. A}} {{#2}}, (#4) #3}

\newcommand{\sci}[4]{{{Science}} {{#2}}, (#4) #3}

\begin{document}
\begin{frontmatter}

\title{Towards the understanding of Decoherence on Ion Traps}
\author{M.O. Terra Cunha}
\corauth{Corresponding author}
\ead{tcunha@mat.ufmg.br}
\address{Departamento de Matem\'atica, ICEx, UFMG, CP 702, Belo Horizonte,
30123-970, Brazil}

\author{M.C. Nemes}
\address{Departamento de F\'{\i}sica, ICEx, UFMG, CP 702, Belo Horizonte,
30123-970, Brazil}

\begin{abstract}
Two mechanisms of decoherence in ion traps are studied,
specially related to the experiment [Kielpinski {\it{et al.}}, {{Science}}
 {{291}} (2001) 1013]. Statistical hypothesis are made
about unknown variables and the expected behaviour of the visibility of
the best experimental pattern is calculated for each mechanism. Data from the 
experiment are analyzed and show to be insufficient to distinguish between them.
We suggest improvements which can do this with slight modifications in the
present facilities.
\end{abstract}

\end{frontmatter}

Ions traps are physical systems over which experimentalists have a large amount
of precise control of the relevant degrees of freedom. This naturally makes
these systems good candidates for exploring quantum phenomena and even for small
scale quantum computation. Decoherence is the major barrier to be overtaken. One
way of understanding decoherence is as a kind of {\emph{quantum
noise}}\cite{NC}. Its consequence is the transformation of pure states, which
can be described by a state vector $\ket{\psi}$, into mixed states to which only
density operator description applies. Naturally, there is a huge set of
decoherence sources, from spurious electromagnetic fields originated in the lab,
passing through small oscillations in parameter controls, up to background
radiation and gravitational waves or whatever other kind of perturbation which
forbids the system to be completely isolated\cite{Dec}. Usually they are all
considered as an {\emph{environment}} to which the system is coupled.

In this work we give special attention to a paper by the Nist
group\cite{Kiepetal01}, in which they showed how to produce a decoherence
``free'' qubit from a pair of ions trapped together. The idea of decoherence
free subspaces can be classically understood, despite the fact of being a
quantum phenomenon: if two classical moving particles are subjected to exactly
the same noise, their relative motion is unaffected. In fact, if one can
determine the major source of decoherence, and find two state vectors which are
affected in the same way by free Hamiltonian and this decoherence source, any
quantum information encoded in the subspace generated by these two vectors will
be preserved from this decoherence source. As a cat and rat dispute, the other
sources of decoherence will prevent this subspace to be completely decoherence
free, but decoherence times will be substantially raised. That is what was
obtained in two different situations in the cited experiment.

We will start our analysis by considering that the major source of decoherence
are oscillations in the magnetic fields used to split hyperfine structure of
$^9{\mathrm{Be}}^+$ ions. More generally speaking, in situations like this, one
can just consider the system to be subjected to a Hamiltonian $\op{H}\de{\xi}$,
which depends on a stochastic parameter $\xi$. Stochastic Hamiltonians generate
stochastic evolutions, so pure states evolve to mixed ensembles in a quantum
state diffusion picture\cite{QSD}. If we consider a situation in which only the
magnitude of the field is subjected to oscillations, then the eigenvectors of
$\op{H}\de{\xi}$ do not depend on $\xi$, but their eigenvalues do. Let us
consider a model like this for the ions: we are interested only in two internal
states of each ion, so we consider the stochastic Hamiltonian to be ($\hbar =
1$)
\begin{equation} \op{H}\de{\xi} =  \frac{\omega _1\de{\xi}}{2}
\op{\sigma}_{z1} + \frac{\omega _2\de{\xi}}{2}  \op{\sigma}_{z2}, \label{Hamxi}
\end{equation}
which can be rewritten as
\begin{equation} \op{H}\de{\xi} =
\frac{\omega _m\de{\xi}}{2}  \DE{\op{\sigma}_{z1} + \op{\sigma}_{z2}} +
\frac{\omega _d\de{\xi}}{2} \DE{\op{\sigma}_{z1} - \op{\sigma}_{z2}},
\label{Hamxim} \end{equation}
where $\omega _m = \de{\omega _1 + \omega _2}/2$
and $\omega _d = \de{\omega _1 - \omega _2}/2$. The form (\ref{Hamxim}) makes
evident the structure of the Hamiltonians $\op{H}\de{\xi}$, including their
dependence on $\xi$. In the ideal case, $\omega _1 = \omega _2$, we get a doubly
degenerate level with energy $E=0$, and two other levels with energies $\pm
\omega _m$. If the perturbations keep $\omega _1 = \omega _2$ we have a
bidimensional decoherence free subspace generated by $\DE{\ket{\up \down},
\ket{\down \up}}$. So the situation close to DFS is to have $\abs{\omega _d} \ll
\abs{\omega _m}$, and to consider initial states in this
subspace. In this subspace, the term proportional to $\omega _m$ is ineffective,
and decoherence is only originated by the variations in $\omega _d$ (\ie only
relative variations are important, because they affect Born frequencies).

Now we address the question: how does a state in a decoherence ``free'' subspace
decohere? Let us consider an initially pure state
\begin{equation}
\ket{\psi \de{0}} = \alpha \ket{\up \down} + \beta \ket{\down \up}.
\label{initstate}
\end{equation}
For a fixed value $\xi$, the Hamiltonian (\ref{Hamxim}) implies that at time $t$ one has
\begin{equation}
\ket{\psi \de{t}}_{\xi} = \alpha e^{-i\omega _d\de{\xi}t} \ket{\up \down} +
\beta e^{i\omega _d\de{\xi}t} \ket{\down \up}. \label{xistatet}
\end{equation}
It is interesting to stress that in this model the vector state (\ref{xistatet})
always belongs to the decoherence ``free'' subspace, but due to its dependence
on $\xi$, decoherence emerges. Writing the state (\ref{xistatet}) as a density
operator and taking the ensemble average over the stochastic parameter $\xi$,
with weights $p\de{\xi}$, one gets
\begin{equation}
\op{\rho}\de{t} = \De{
\begin{array}{cc} \abs{\alpha}^2 & k^*\alpha ^*\beta \\
k\alpha \beta ^* & \abs{\beta}^2
\end{array}},
\label{rhok}
\end{equation}
where we defined the parameter
\begin{equation}
k = \mean{e^{i\omega _d\de{\xi}t}} = \int _Q e^{i\omega _d\de{\xi}t} p\de{\xi}d\xi .
\label{defk}
\end{equation}

Let us stop for a moment to discuss states like (\ref{rhok}). A Bloch vector
picture is simple, specially in cylindrical coordinates: $\abs{\alpha}^2 -
\abs{\beta}^2$ determines $z$ coordinate, and $2\abs{k\alpha \beta ^*}$ is the
cylindrical radius (\ie the distance to the $z$ axis). The most interesting case
is $\abs{\alpha}^2 = \abs{\beta}^2 = \frac{1}{2}$ (which corresponds to the
experimental case), when $\abs{k}$ directly gives the norm of Bloch vector,
which can be interpreted as a direct measure of the purity of the state. In
practice, the norm of a Bloch vector can be associated to the visibility of the
best interferometer prepared with the state $\op{\rho}$. In interferometric
experiments one naturally searches for the largest visibility\cite{CN02},
therefore we shall focus our attention on the factor $k$.

Now let us discuss the factor $k$ of the model here presented, eq.\
(\ref{defk}). As the only influence of $\xi$ we are considering is on the
frequency $\omega _d \de{\xi}$, we can consider as stochastic parameter a
frequency $\nu$ itself, with an ensemble weight $p\de{\nu}$. So we have
\begin{equation} k = \int _{-\infty}^{\infty} e^{i\nu t} p\de{\nu}d\nu ,
\label{knu} \end{equation} where one can recognize the structure of Fourier
transform of the stochastic weight $p\de{\nu}$, so common in optics\cite{FO}. As
$k$ determines the visibility of the best interferometer prepared with
state $\op{\rho}$, the data of ref.\ \cite{Kiepetal01} must fit this
expression for $\abs{k}$. An interesting improvement on the experiment
would be to independently monitor the magnetic field, and try to
compare the fringe visibility and the Fourier transform of the
statistical variations of the field. This would decide whether
fluctuations of the magnetic field are the major decoherence source or
not. As these fluctuations are not available, we can only speculate about them.
One natural hypothesis is of Gaussian distribution. This would imply a Gaussian
decay of $\abs{k}$ with time $t$. Other distributions can also appear, but we
have no a priori reasons for treating them. One last example is to suppose
exponential decay for the visibility, what in this model would be consistent
with Lorentzian stochastic weight
\begin{equation}
 p\de{\nu} \propto
\frac{\Gamma}{\de{\nu - \nu _o}^2+ \Gamma ^2} .
\label{pLorentz}
\end{equation}

In the cited experiment\cite{Kiepetal01}, one state in the decoherence
``free'' subspace (DFS state) and one out of it (test state) were submitted to
two distinct situations each: autonomous evolution, just subjected to natural
noise, and an engineered reservoir, consisting of a dissonant laser with random
intensity. The first situation is supposed to be well modeled by the above
discussion on variable magnetic field, but the second one deserves special
attention. Experimental values allow one to consider only decoherence sources
related to the presence of the dissonant laser. The most simple way to do it
would be to consider the dispersive approximation to Jaynes-Cummings model for
the interaction among one field mode and two two-level ions: \begin{equation}
\op{H}_{dJC} = \op{H}_o + \op{H}_{int}, \end{equation} where \begin{equation}
\op{H}_{o} = \frac{\omega}{2}  \op{J}_{z}+ \omega _f\op{a}^{\dagger}\op{a},
\label{Ho} \end{equation} where $\op{J} = \op{\sigma}_1 + \op{\sigma}_2$, and
\begin{eqnarray} \op{H}_{int} &=& g\de{\op{J}_+\op{a} +
\op{J}_-\op{a}^{\dagger}} \nonumber \\ &\approx & \Omega
\DE{\op{a}\op{a}^{\dagger}\de{\ket{\up _1}\bra{\up       _1} + \ket{\up
_2}\bra{\up _2}} -   \op{a}^{\dagger}\op{a}\de{\ket{\down _1}\bra{\down _1} +
\ket{\down       _2}\bra{\down _2}}}, \label{dispapprox}
\end{eqnarray}
where $\Omega = \frac{g^2}{2\delta}$, $g$ is a dipolar coupling
constant, $\delta = \frac{1}{2}\de{\omega - \omega _f}$ is the
detuning, and dispersive approximation was applied, under the
condition $\delta ^2 \gg g^2\de{n+1}$, $n$ the number of photons.
However, this can not mimic the experimental situation, unless we
include somehow the randomization of the laser field. This can be
done by considering an environment coupled to this field. As usual,
let us consider a huge set of harmonic oscillators and bilinear
coupling as a model to the environment. In this sense, the full
Hamiltonian we must consider is
\begin{equation}
\op{H} = \op{H}_{dJC} + \sum \omega _l \op{b}_l^{\dagger}\op{b}_l +
\sum g_l\de{\op{a}^{\dagger}\op{b}_l + \op{b}_l^{\dagger}\op{a}}.
\label{modeldissip}
\end{equation}
The laser-environment part of this Hamiltonian can be rewritten in
terms of normal modes, and this formally shows the equivalence between two
ions coupled to a laser coupled to an environment and the same two
ions directly coupled to a reservoir. To go further one needs to make
some hypothesis on the coupling of the laser to the environment. White
noise is a natural choice, and a Wigner-Weisskopf like
approximation\cite{Lou} implies the form of the coupling between ions
and environmental normal modes. This is once again consistent with
exponential decay of fringe contrast in interferometry.

So, we considered two distinct sources of decoherence for a system like
the one worked out in ref. \cite{Kiepetal01}: one given by stochastic
variations of the energy levels (probably due to variations of the applied
magnetic field), and other given by the direct coupling to a
``classical'' field mode (``classical'' in the sense of being strongly
coupled  to an environment).  For both, we made statistically natural
hypothesis, namely  Gaussian fluctuations and white noise. The first
model naturally gives Gaussian decay for the fringe contrast while the
second implies exponential decay. We therefore strongly believe that when the
laser is off, ``natural'' decoherence will fit a Gaussian, while with the laser
on, ``engineered'' decoherence will be exponential. We take the data from the
experiment, and used least square methods to obtain the best Gaussian and the
best exponential to each set of data, and also used a kind of sieve to try to
tell us if the points fit better a Gaussian or an exponential. We now discuss in
detail the methods we used and the results we got.

In fact, the function which we statistically treated was $F\de{t} = \ln
V\de{t}$,
where $V$ is the visibility of the fringe pattern. For each set of
data we search for the least square curves of the form
\begin{eqnarray}
F &=& at + b, \nonumber \\
F &=& At^2 + B,
\end{eqnarray}
for which is important to say that both have the same number of
parameters, what make the comparison fair. For each set of data, and
each fitting family, we
accumulate the vertical distances between the measured points and the
best fitting curve and compare: for each set of data, the one which
accumulate less distance is considered the best curve. For each
initial state and laser situation the figure shows the experimental points, the
best Gaussian (in fact, the best parabola for the logarithm) and the best
exponential (resp. linear function). The accumulated distance is given in
captions. The worked data are not enough to corroborate or deny our
pre-conclusions about the form the coherence decays in each case, although a
few more points could do the job. Also, as pointed out before, independently
recording the field fluctuations and comparing its line shape to the visibility
(with the laser off) could test the usually accepted viewpoint that the most
important decoherence source are these field fluctuations. We expect that this
work can stimulate more detailed experiments towards a better understanding of
the mechanisms of decoherence on ion traps.

\section*{Acknowledgements}
The authors acknowledges enlightening discussions with Dr. Ruynet L. de Matos
Filho,  and thank Drs. D. Kielpinski and D. Wineland for sending to us the data
from their experiments. This work is supported by the Millenium Institute on
Quantum Information, from Brazilian agency CNPq.

\section*{Figure captions}

Figure 1:
For autonomous decoherence (laser off), the points and error bars are
experimental\cite{Kiepetal01}, the curves are least square fits: {\bf{(a)}} DFS
state, $y = -.803 -.00224 t$, with accumulated square distance (asd) $.0095$,
and $y = -.997 -.393 \times 10^{-5} t^2$, with asd $.062$; {\bf{(b)}} Test
state, $y = -.109 -.00883 t$, with asd $.037$, and $y = -.394 -.391 \times
10^{-4} t^2$, with asd $.0040$.

Figure 2:
The same for engineered decoherence (laser on): {\bf{(a)}} DFS state,
$y = -.874 - .00330 t$, with asd $.0084$, and $y = -.884 -.159 \times 10^{-3}
t^2$, with asd $.0083$; {\bf{(b)}} Test state, $y = -.174 -.175 t$, with asd
$.084$, and $y = -.0110 - .581 t^2$, with asd $.16$.

\end{document}